\documentclass[ twocolumn]{aastex631}

\newcommand\ms{$\textrm{m~s}^{-1}$}

\newcommand\teff{T$_{\rm{eff}}$}
\newcommand\vsini{$v$sin$i$}

\newcommand\earthinsol{$S_{\oplus}$}
\newcommand\earthradius{$R_{\oplus}$}
\newcommand\solmass{$M_{\odot}$}

\newcommand\surveyname{\textit{Searching for GEMS}}

\hypersetup{linkcolor=violet,citecolor=purple}

\graphicspath{{./}{figures/}}

\received{September 16, 2024}
\revised{November 17, 2024}
\accepted{November 25, 2024}
\submitjournal{ApJ}

\begin{document}

\title{Transiting Jupiters around M-dwarfs have similar masses to FGK warm-Jupiters}

\author[0000-0001-8401-4300]{Shubham Kanodia}
\affiliation{Earth and Planets Laboratory, Carnegie Science, 5241 Broad Branch Road, NW, Washington, DC 20015, USA}

\shortauthors{Kanodia 2024}

\correspondingauthor{Shubham Kanodia}
\email{skanodia@carnegiescience.edu}

\begin{abstract}
This paper presents a comparative analysis of the bulk properties (mass and radius) of transiting giant planets ($\gtrsim$ 8\earthradius) orbiting FGKM stars. Our findings suggest that the average mass of M-dwarf Jupiters is lower than that of their solar-type counterparts, primarily due to the scarcity of super-Jupiters ( $\gtrsim$ 2 $M_J$) around M-dwarfs. However, when super-Jupiters are excluded from the analysis, we observe a striking similarity in the average masses of M-dwarf and FGK warm-Jupiters. We propose that these trends can be explained by a minimum disk dust mass threshold required for Jovian formation through core accretion, which is likely to be satisfied more often around higher mass stars. This simplistic explanation suggests that the disk mass has more of an influence on giant planet formation than other factors such as the host star mass, formation location, metallicity, radiation environment, etc., and also accounts for the lower occurrence of giant planets around M-dwarf stars.  Additionally, we explore the possibility of an abrupt transition in the ratio of super-Jupiters to Jupiters around F-type stars at the Kraft break, which could be a product of \vsini{} related detection biases, but requires additional data from an unbiased sample with published non-detections to confirm. Overall, our results provide valuable insights into the formation and evolution of giant exoplanets across a diverse range of stellar environments.

\end{abstract}

\keywords{}

\section{Introduction} \label{sec:intro}
The launch of NASA's Transiting Exoplanet Survey Satellite \citep[TESS;][]{ricker_transiting_2014} has enabled the discovery of over two dozen transiting Giant Exoplanets around M-dwarf Stars \citep[GEMS, $R_p \gtrsim$ 8 \earthradius;][and others]{triaud_m_2023, kanodia_searching_2024}. These transiting planets, despite typically being in short-period orbits ($\lesssim$ 5 days), are (i) much cooler ($<$ 800 K) than standard hot-Jupiters and (ii) found at larger separations to their host stars  (a/R$_*$ $\gtrsim$ 10). 

Recent detections include massive GEMS such as TOI-5205b, TOI-4201b, TIC-46432937b and TOI-2379b, with increasingly higher planet-to-stellar mass ratios \citep[at 0.26\%, 0.39\%, 0.54\% and 0.85\%, respectively][]{kanodia_toi-5205b_2023, delamer_toi-4201_2024, hartman_toi_2024, bryant_toi-2379_2024}. These high mass ratios can be a challenge to explain with the standard core-accretion model of formation given current class II protoplanetary disk samples \citep{manara_demographics_2022}, and the amount of time it is expected to take for a M-dwarf disk to accrete a massive enough core to initiate the runaway processes of gaseous accretion \citep{laughlin_core_2004, burn_new_2021}. While such high mass-ratio planets are not a typical outcome of planet formation, by now it is also well established that the class II disk masses are likely underestimated by almost an order of magnitude \citep{xin_measuring_2023, liu_underestimation_2022}, and that the core formation precedes the class II phase \citep{greaves_have_2010, tychoniec_dust_2020, savvidou_there_2024}. Therefore, it is possible for these planets to form through core-accretion in disks that are just massive enough but are not gravitationally unstable and prone to instabilities \citep{boss_giant_1997, boss_rapid_2006}. Gravitational instability has been as a potential alternative to core-accretion for these massive objects, which is particularly attractive for M-dwarf hosts given the challenges of insufficient disk masses and long formation timescales with core accretion \citep{boss_rapid_2006, boss_forming_2023}. However, these higher mass-ratio planets are likely few and far between, potentially being outliers in the transiting GEMS sample, which are already a rarity \citep{gan_occurrence_2023, bryant_occurrence_2023}.

\begin{figure*}[!t]
\centering
\begin{tabular}{ccc}
 \includegraphics[scale=0.33,clip]{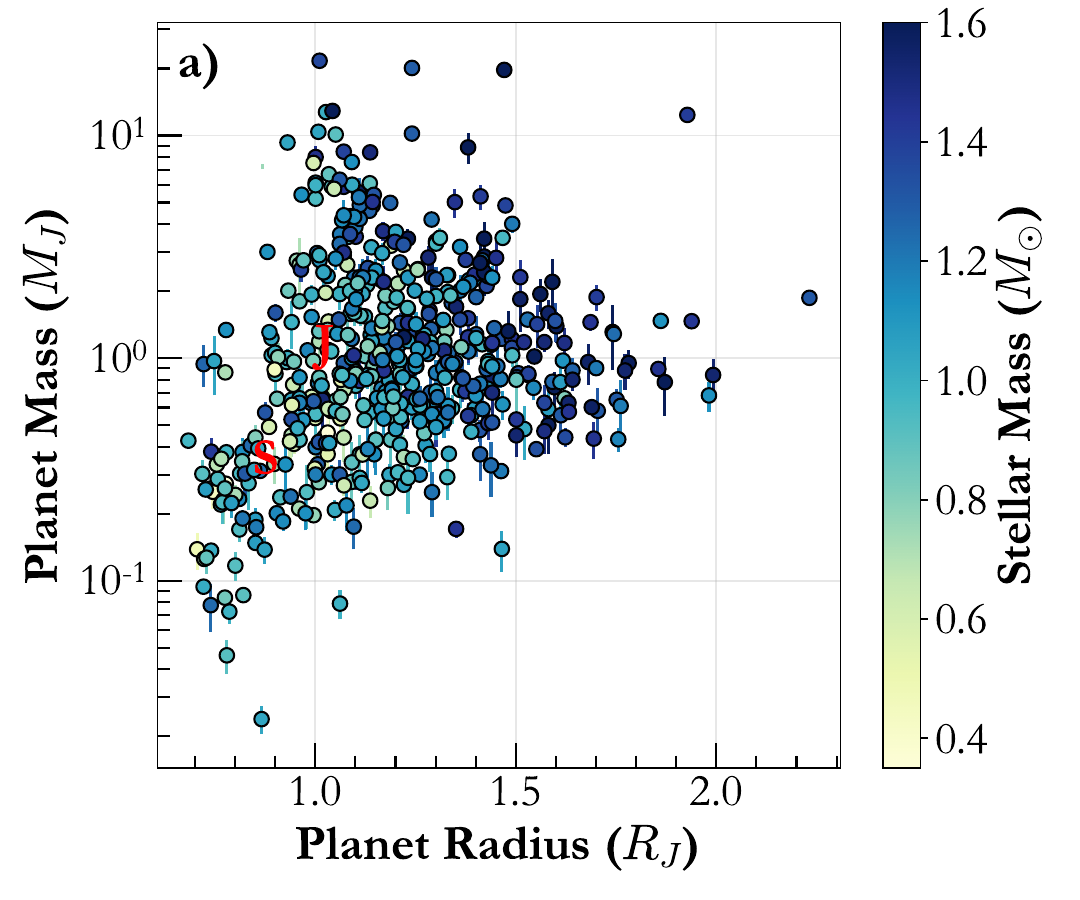} &
 \includegraphics[scale=0.33,clip]{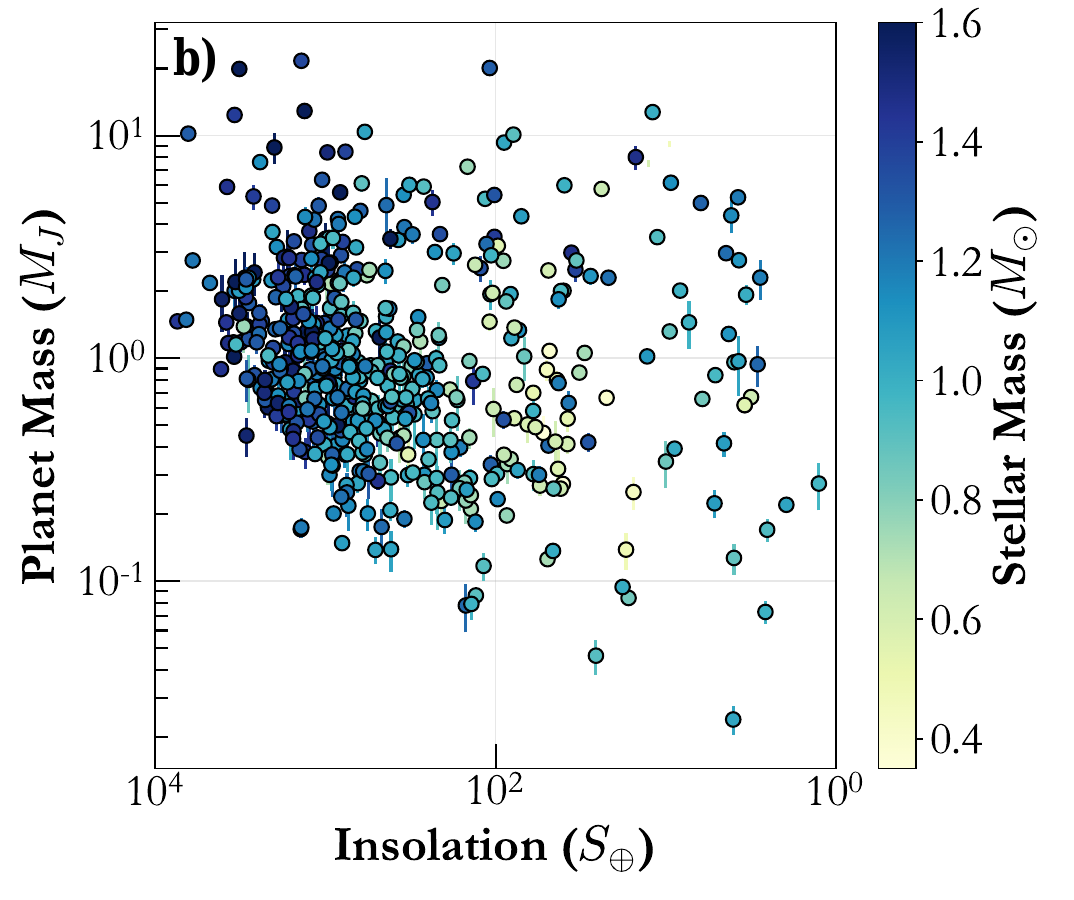} & 
 \includegraphics[scale=0.33,clip]{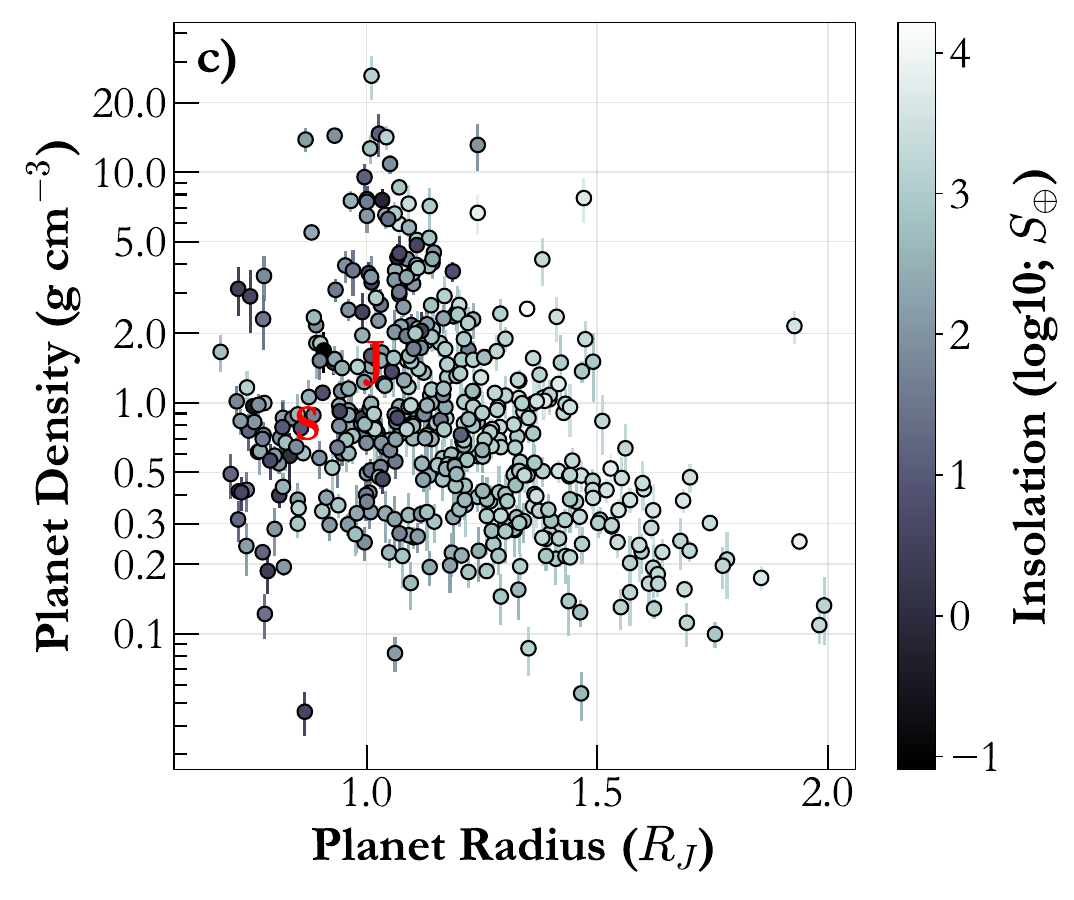} 
 \end{tabular}
\begin{tabular}{ccc}
 \includegraphics[scale=0.35,clip]{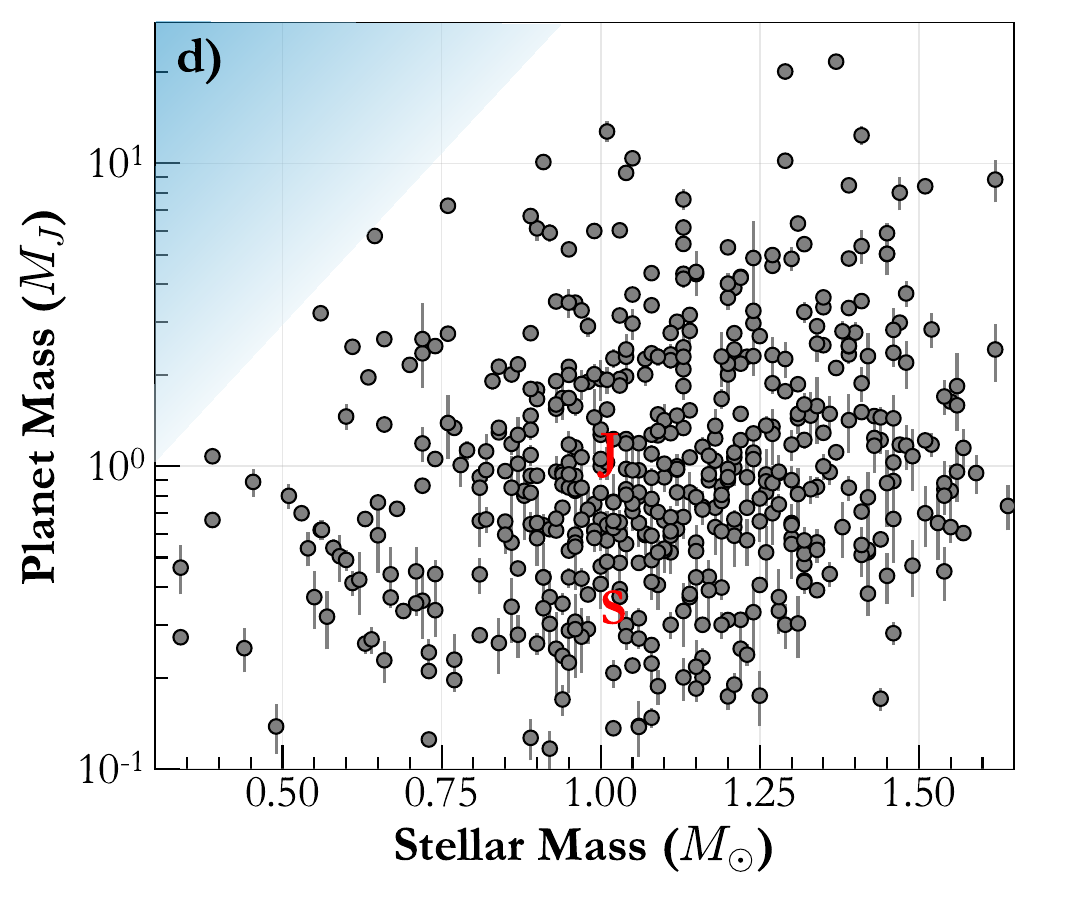} &
 \includegraphics[scale=0.35,clip]{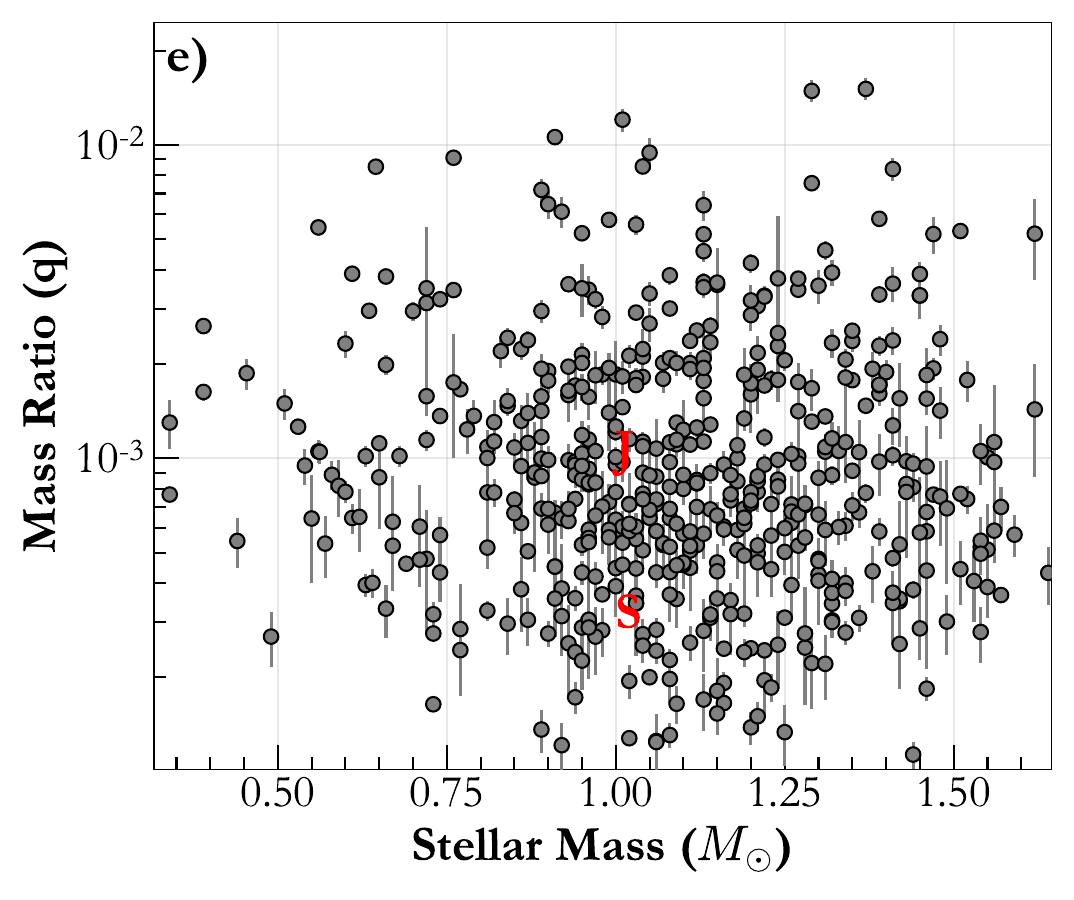}
\end{tabular}    
\caption{\small  All the planets in our sample in \textbf{a)} planet mass - radius space colour coded by stellar mass, \textbf{b)} planet mass - insolation colour coded by stellar mass, \textbf{c)} planet density - radius colour coded by insolation,  \textbf{d)} planet mass as a function of stellar mass, and  \textbf{e)} planet-to-stellar mass ratio as a function of stellar mass. We indicate Jupiter and Saturn in red with `J' and `S' respectively. We also include a shaded blue region in panel \textbf{d} showing a region devoid of massive planets, which suggests a correlation with host star masses. The same zone is not as apparent in the mass-ratio panel \textbf{e}.}\label{fig:scatter}
\end{figure*}

To better understand the \textit{average} transiting GEMS sample, we have started the \surveyname{} survey to increase the number of well-characterized transiting GEMS with mass measurements to $\sim$ 40 planets for robust statistical inference. Specifically, to address whether the difficulty in forming GEMS in this mass starved regime through core-accretion manifests as a systematic difference in their bulk-properties as a function of stellar mass.

With half the survey data already collected and when combined with detections from the community, we start to see interesting trends in the dataset that we discuss in this manuscript. Specifically, in this work we are interested in addressing the question, Are the properties of giant planets different around stars of different masses? This question requires studying the normalized conditional probabilities of planets, not their absolute occurrence or frequency. We describe our sample in Section \ref{sec:sample}, and the analysis in Section \ref{sec:analysis}. We then discuss the implications of our inference in Section \ref{sec:discussion}, before concluding in Section \ref{sec:conclusion}.

\section{Sample}\label{sec:sample}
\subsection{Selection}
We query the Planetary Systems Table from the NASA Exoplanet Archive \citep{ps}\footnote{Accessed on 2024 September 11.} for host stars with \teff{} $<$ 7200 K to include FGKM stars with radii measurements $>$ 5$\sigma$ and mass measurements at $>$ 3$\sigma$. We then limit the sample to planets with radius $\gtrsim$ 8 \earthradius, and exclude systems without listed stellar or planetary mass measurements and uncertainties. To this we add recent GEMS systems that are not yet on the archive, including those published by the community --- TOI-2379b and TOI-2384b \citep{bryant_toi-2379_2024} --- and by us --- K2-419Ab, TOI-6034b, \citep{kanodia_searching_2024-1}, TOI-6383Ab \citep{bernabo_searching_2024} and TOI-5688Ab \citep{reji_searching_2024}, which results in 548 planets.  

Our sample is shown in \autoref{fig:scatter}, in planetary mass, planetary radius and stellar mass space. We note a few salient features of the planetary mass--radius plane: (i) The inflated giant planets ($R_p >> 1 R_J$) are preferentially seen around higher-mass host stars, and higher insolation fluxes \citep{guillot_evolution_2002, weiss_mass_2013, thorngren_bayesian_2018}, (ii) these inflated radii are generally not seen for the more massive giant planets ($>$ 2 $M_J$) with higher gravity, (iii) there are not  many hot and inflated Saturns ($R_p >> 1 R_J$, $M_p \sim 0.3~M_J$, $S_p >~200$\earthinsol), which has previously also been noted as evidence of runaway mass-loss \citep{thorngren_bayesian_2018, thorngren_removal_2023}, and (iv) in the planet mass--stellar mass plane, we note a tentative trend seen between maximum planetary mass and host-star mass that is marked in blue in Figure~\ref{fig:scatter}d.

\subsection{Observational Biases}\label{sec:biases}
In this analysis, we do not account for survey detection statistics and observational biases from the different surveys (\textit{Kepler}, TESS, and ground-based) while comparing planet samples across stellar mass. Despite this, our inferences should be unaffected because while the occurrence rate estimates across planet radius -- orbital period are sensitive to survey completeness, these estimates are agnostic of planet masses in the regime of interest in this work ($\gtrsim$8~$R_{\oplus}$). Furthermore, while there are clear stellar mass dependent trends in the number of giant planets around different host stars, the analysis performed here is restricted to the properties of Jovian-sized planet samples and not their number, i.e., we are not comparing histograms of planet properties but their conditional probabilities. And while the number of planets in a given $R_p, M_*$ bin affect the precision of the results, fewer planets should not affect the accuracy of the recovered normalized model posteriors.

Biasing the inferences presented in this manuscript would require systematics that are conditional --- $M_p| R_p, M_*$ --- and affect planet masses in M-dwarf and FGK samples differently. We do not expect measurement biases in giant planet masses since such planets in a median FGK system in our sample ($V_{mag}$ $\sim$ 12.2, orbital period $\sim$ 3.8 days; \autoref{fig:scatter}) would have an RV semi-amplitude  $\gtrsim$ 50 \ms{} for planets $\gtrsim$ 0.4 $M_J$, which should be measurable by most RV spectrographs. In other words, a median giant planet in our sample should have a mass ($\sim$ 0.4 $M_J$) that is equally detectable irrespective of an M-dwarf vs FGK host star.

\section{Analysis}\label{sec:analysis}
\subsection{Trends in planet mass}\label{sec:analysismasses}

One of the primary goals of the \surveyname{} survey \citep{kanodia_searching_2024} is to ascertain differences in mass-radius (M-R) relations between giant planets orbiting FGK stars and M-dwarfs. In other words, whether the difficulty in forming GEMS through core-accretion manifests as systematic differences in their bulk properties as a function of stellar mass. We note the potential trend seen in the sample of giant planets in \autoref{fig:scatter}d, where the maximum planetary mass for systems is shown to be correlated with stellar mass. To study this trend and quantify its significance, we utilize the \textit{n}-dimensional sample inference framework \texttt{MRExo} \citep{kanodia_beyond_2023}, which was originally developed for 2-dimensional mass-radius modelling using nonparametric basis functions \citep{ning_predicting_2018, kanodia_mass-radius_2019}, and later expanded to model up to four dimensions simultaneously. \texttt{MRExo} models the \textit{n}-dimensional joint distribution using a convolution of beta-density and normal functions. Regions of parameter space with more planets are represented by beta polynomials (one for each dimension) with a higher joint probability, whereas the normal functions help account for measurement uncertainties. The degree of these betea polynomials (which determines the complexity of the functions) is optimized using 10-fold cross-validation, while the weights are determined through the $MM$ algorithm. The multi-dimensional distribution can then be marginalized to derive normalized conditional distributions, enabling the study of parameter interdependencies without depending on the absolute planet frequency.

\subsubsection{2D: Mass--Stellar Mass ($M_p, M_*$)}\label{sec:2d}

\begin{figure*}[!t]
    \fig{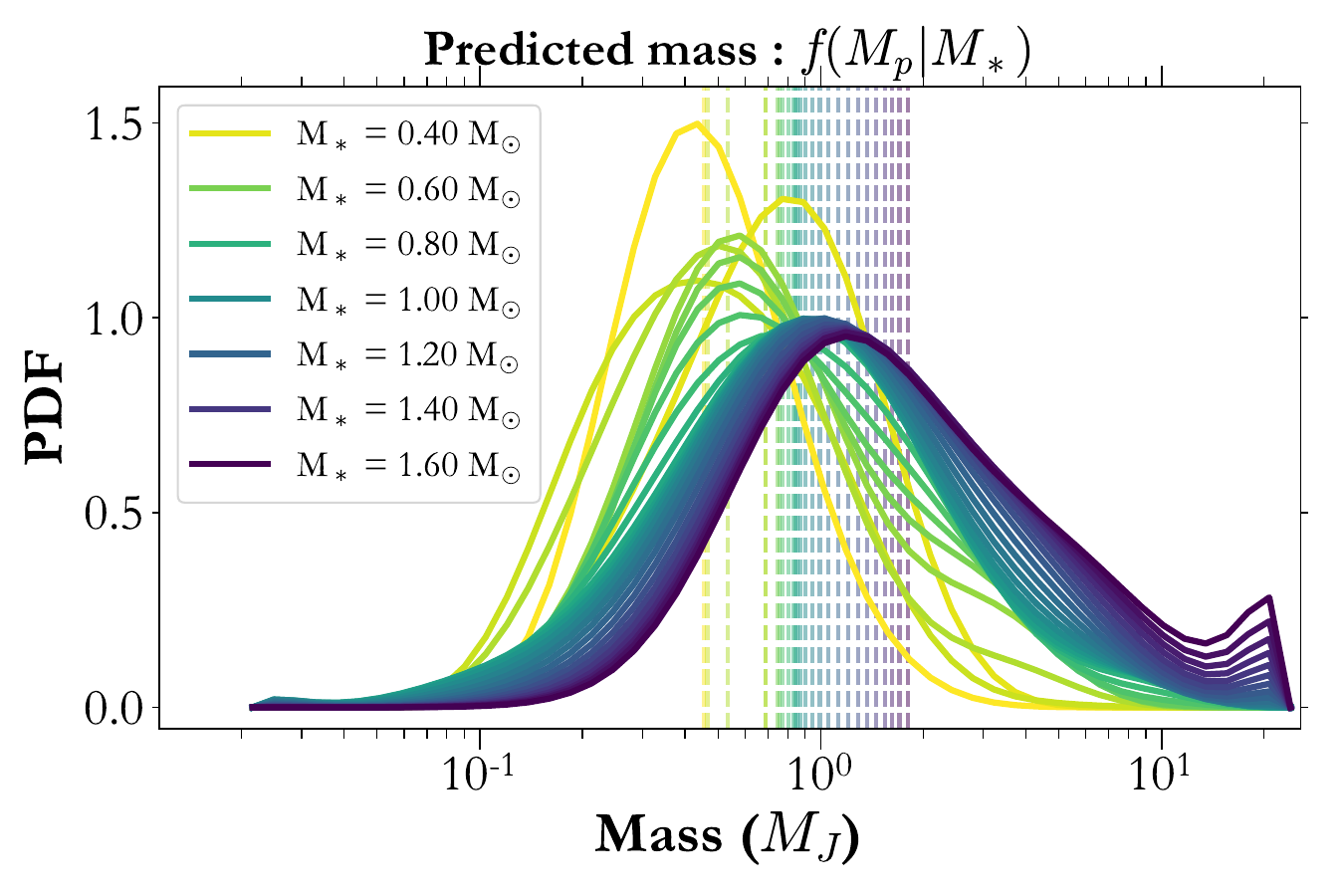}{0.47\textwidth}{}
    \fig{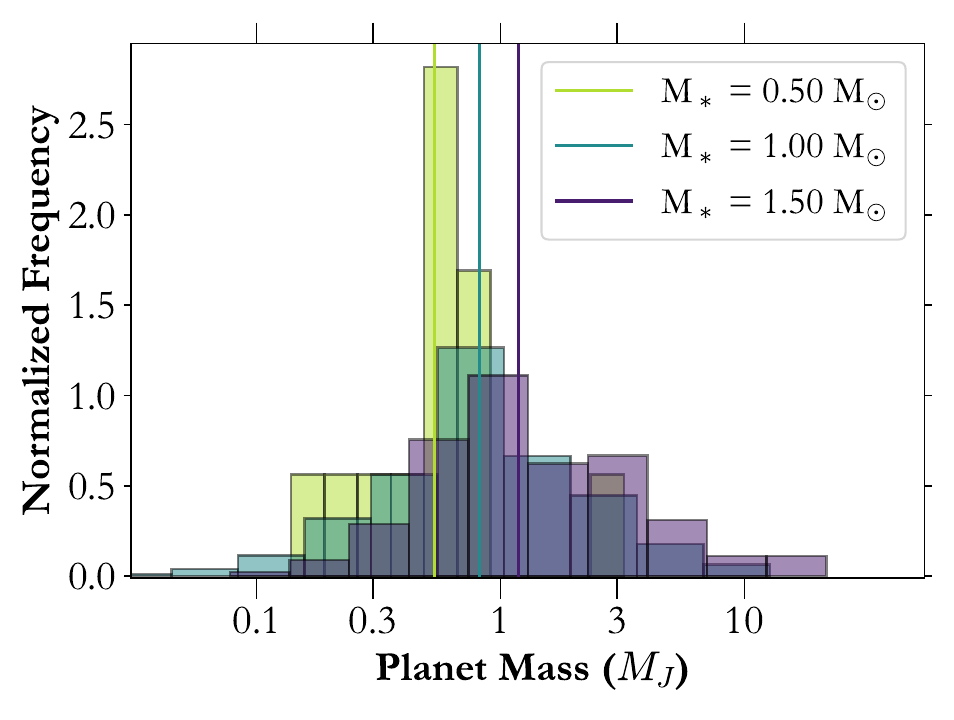}{0.4\textwidth}{}
\caption{\small \textbf{Left)} Conditional distribution of planet masses as a function of stellar mass  --- $f(M_p| M_*)$ --- for our sample, where the dashed lines indicate the expectation value for the distribution. \textbf{Right)} 2D planet mass histograms for different stellar mass bins, where each bin spans $\pm$ 20\% of the nominal stellar mass and the horizontal lines depict the median value of the planet mass distribution.  On average the giant planets around lower mass stars seem to be lower in mass compared to their FGK counterparts. }\label{fig:2D}
\end{figure*}

\begin{figure*}[ht]
\centering
\begin{tabular}{ccc}
 \includegraphics[scale=0.33, trim={0.2cm 0.2cm 2.4cm 0.2cm},clip]{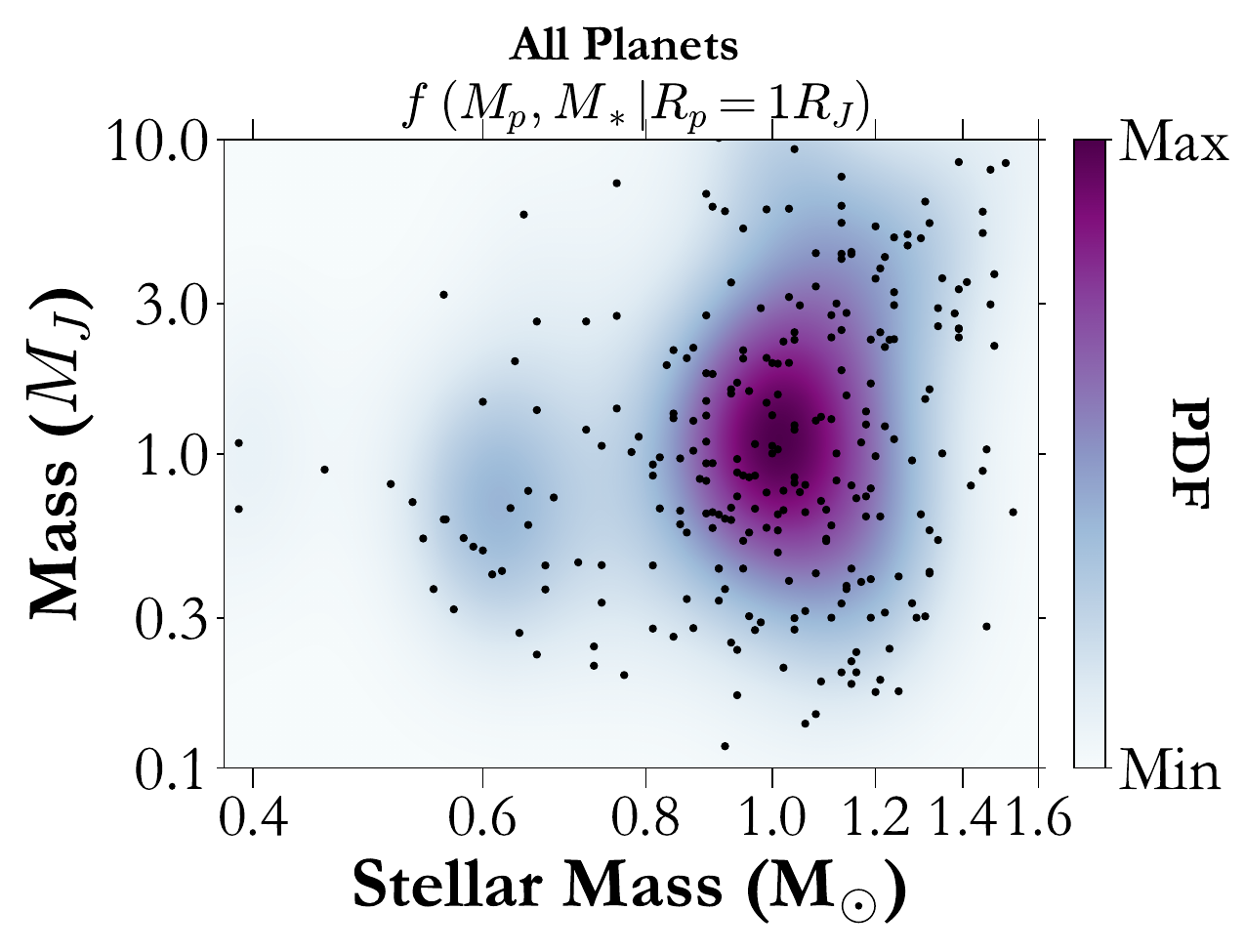} &
 \includegraphics[scale=0.33, trim={3.6cm 0.2cm 2.4cm 0.2cm},clip]{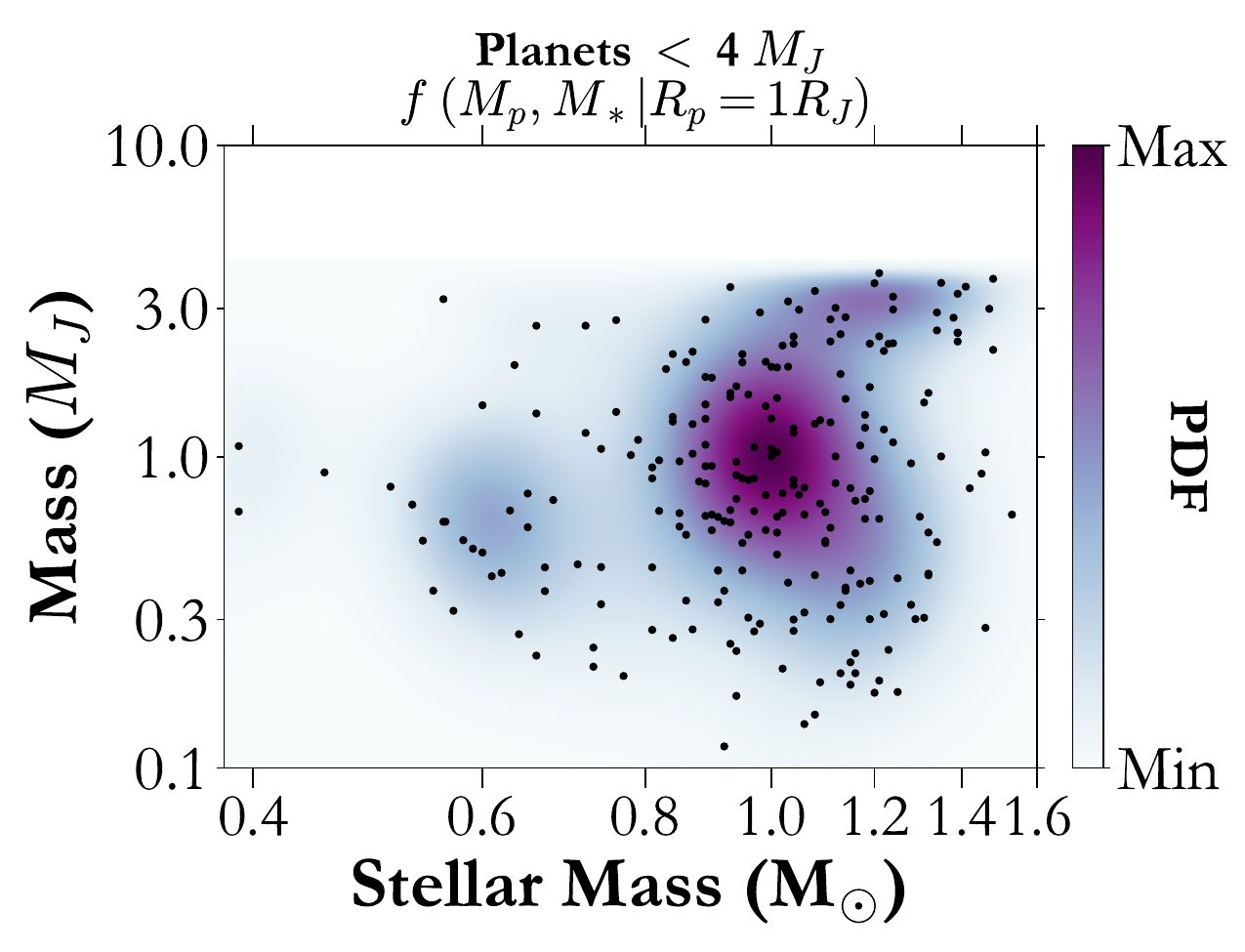} &
 \includegraphics[scale=0.33, trim={3.6cm 0.2cm 0.2cm 0.2cm},clip]{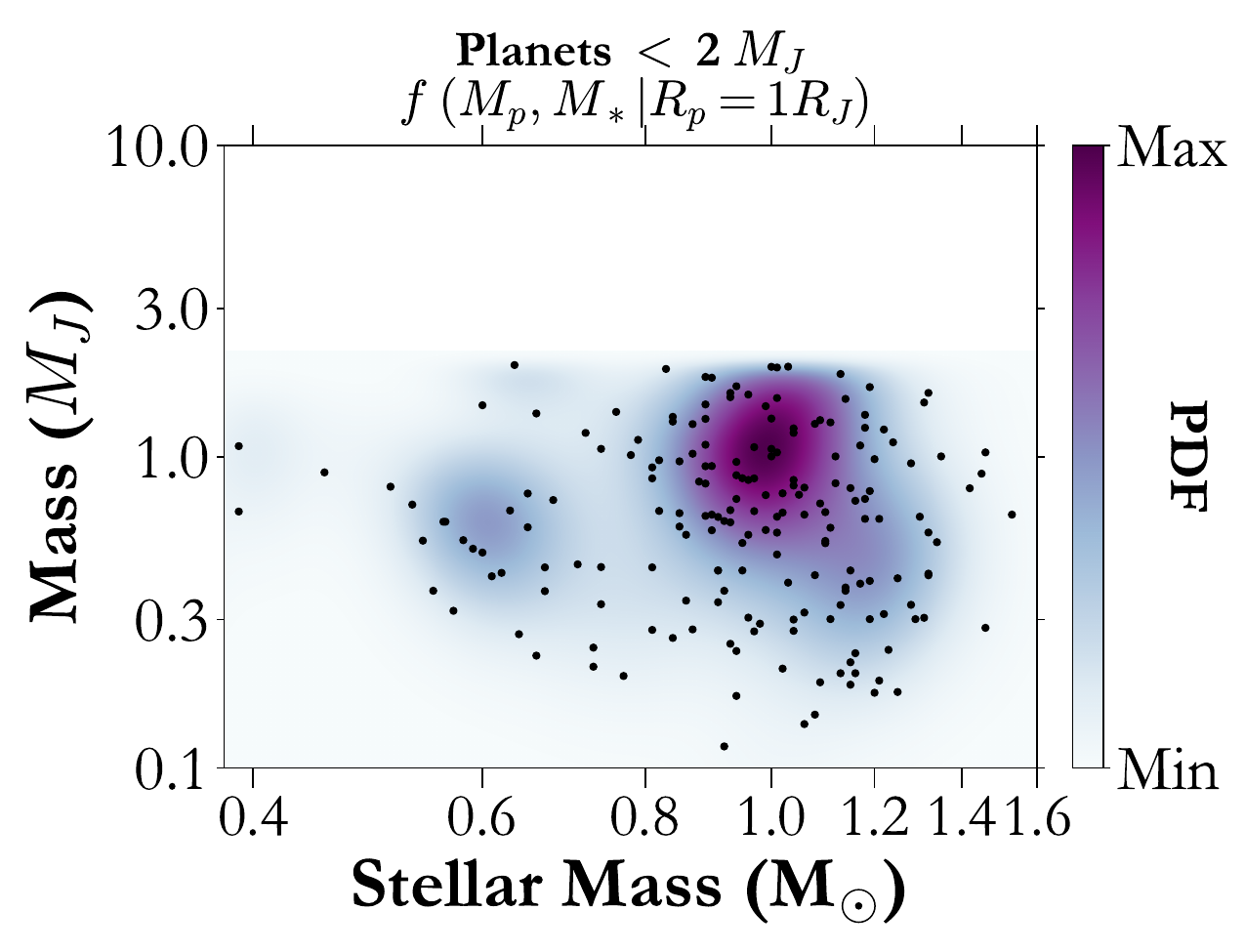}
 \end{tabular}
\begin{tabular}{ccc}
  \includegraphics[scale=0.26, trim={0.2cm 0.2cm 0.2cm 0.2cm},clip]{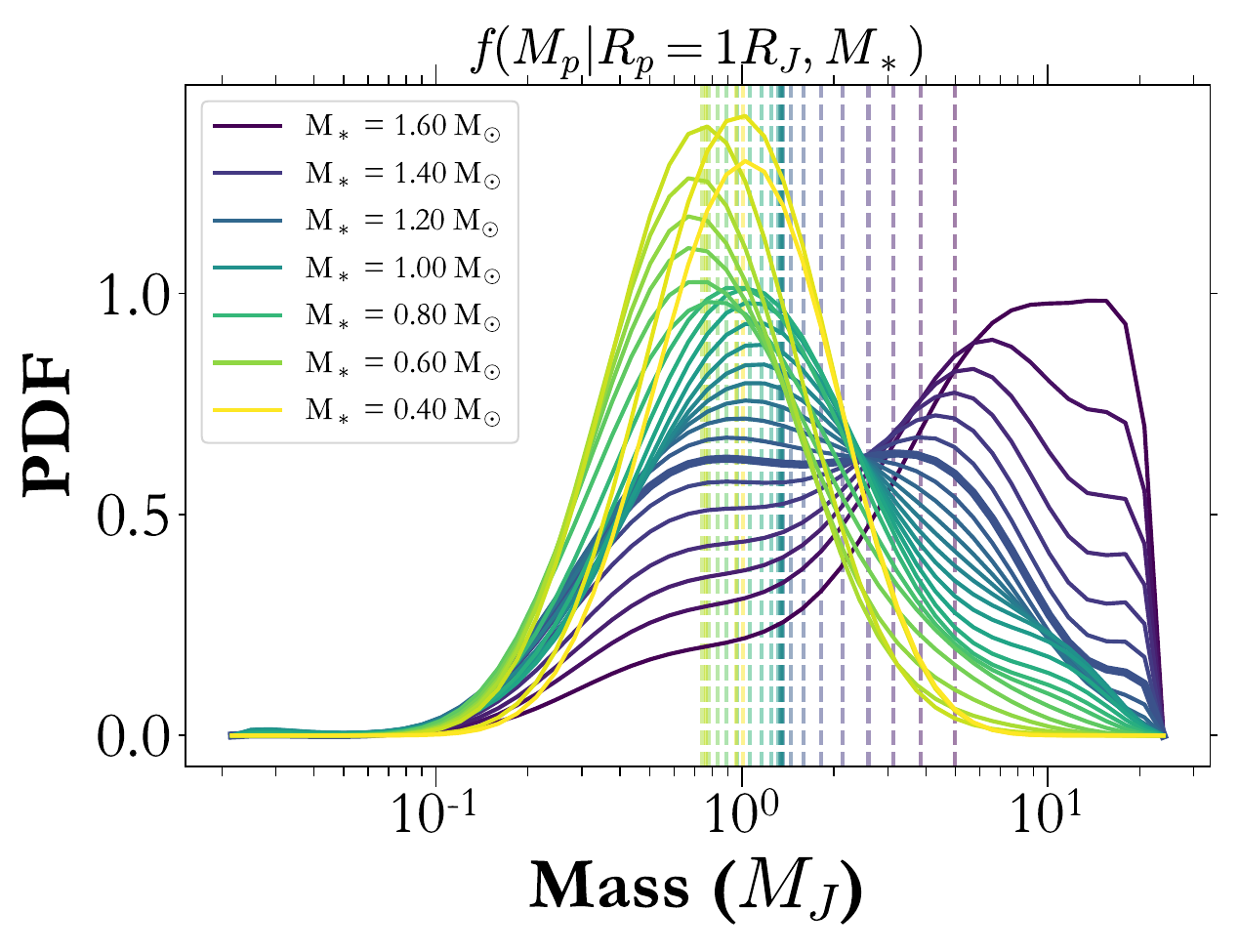} &
 \includegraphics[scale=0.26, trim={0.2cm 0.2cm 0.2cm 0.2cm},clip]{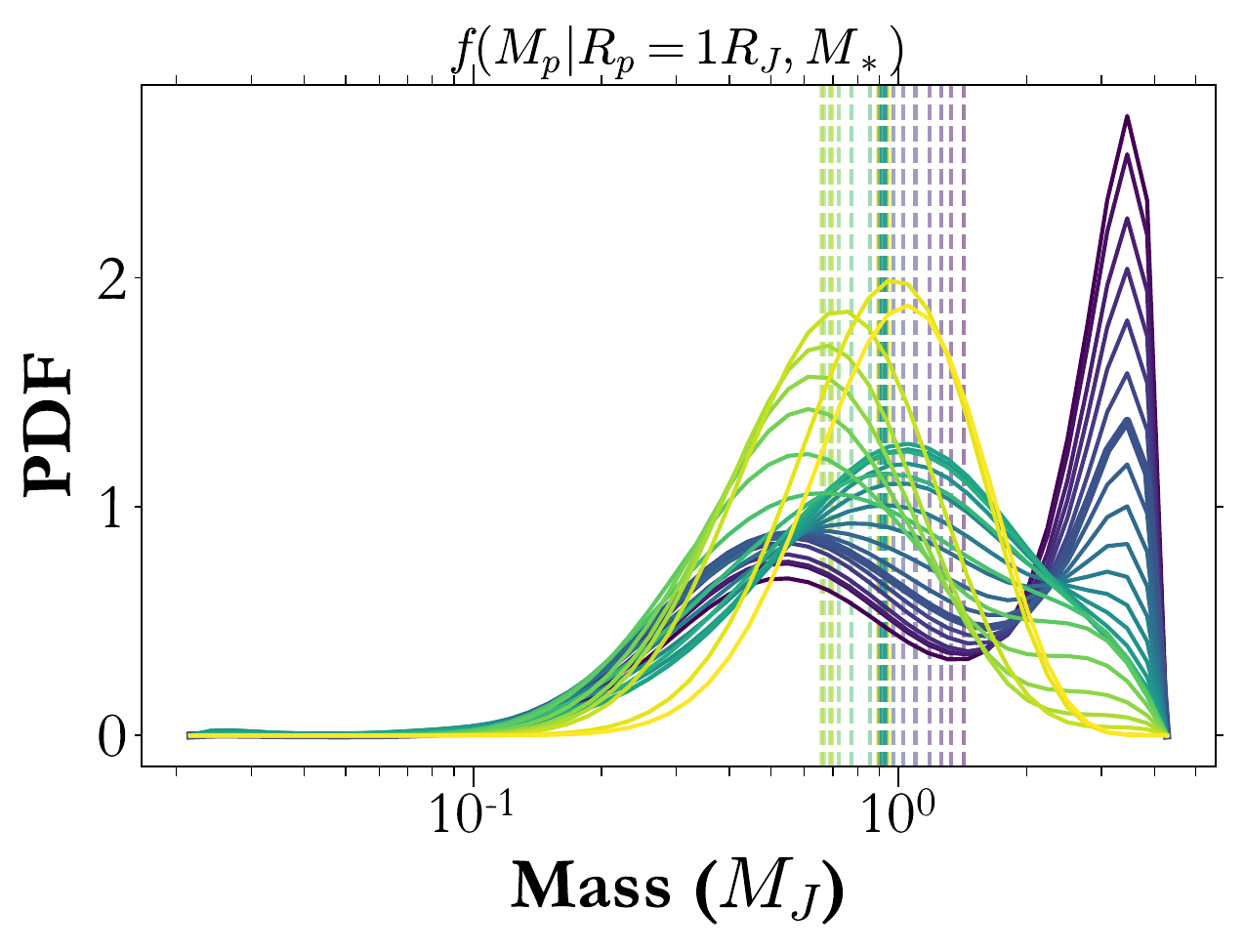} &
 \includegraphics[scale=0.26, trim={0.2cm 0.2cm 0.2cm 0.2cm},clip]{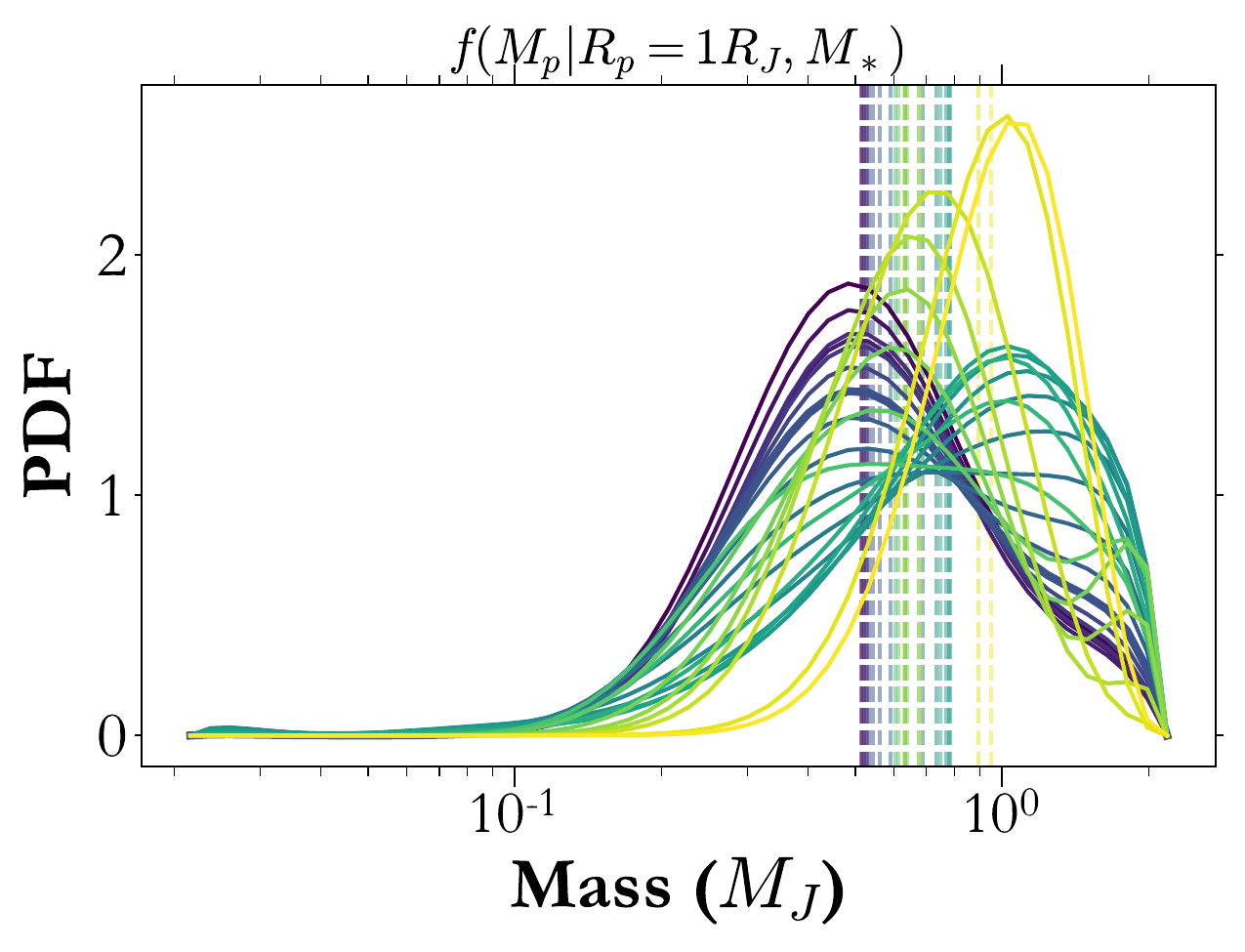} 
\end{tabular}
\caption{The 3D $f(M_p, R_p, M_*)$ fit conditioned on $R_p = 1 R_J$. The top row shows the 2D PDF $f(M_p, M_* | R_p = 1 R_J)$ overlaid with planets between 0.8 -- 1.2 $R_J$. The bottom row shows a 1D PDF for $f(M_p | R_p = 1 R_J, M_*)$ similar to \autoref{fig:2D}. The leftmost column shows results from the full sample, where we see a similar trend as Section \ref{sec:2d} for Jupiter radius planets. The multi-modal nature of the 1D PDF for the more massive host stars seems to suggest two distinct populations of giant planets with Jupiters ($\sim 1 M_J$) and super-Jupiters ( $\sim 10 M_J$). Given this bimodality and to exclude super-Jupiters, we restrict our sample to planets below 4 $M_J$ and repeat the analysis in the middle column, which still seems to suggest a bimodality in the planet distribution. This clearly suggests that a higher mass cut would result in a similar conclusion. In the right-most column we further restrict the sample to only planets $< 2 M_J$ and subsequently find the 1D PDFs across stellar mass to be broadly consistent with each other.}
\label{fig:3DPlMassRadius}
\end{figure*}

We first perform a 2D fit to the planetary mass as a function of stellar mass --- $f(M_p, M_*)$ --- to ascertain the presence of any trend and quantify its significance. Allowing for a maximum of 200 degrees for the beta density functions\footnote{See \cite{ning_predicting_2018} for a review on the implementation of beta density functions for this purpose. Higher degree density functions (or polynomials) allow more complex trends to be modelled.}, the cross-validation technique \citep[described in Section 2.2 in][]{ning_predicting_2018} optimizes for 52, 136 degrees respectively, by maximizing the log-likelihood. This method as shown in \cite{ning_predicting_2018, kanodia_mass-radius_2019, kanodia_beyond_2023} guards against over-fitting by self-regularizing the number and weight of the degrees. For e.g.,  despite having 7072 weights here (136 $\times$ 52), only $\sim 300$ are $> 10^{-3}$. The conditional distribution from this fit --- $f(M_p| M_*)$ --- is shown in \autoref{fig:2D}, alongside histograms of planet mass distributions, where \textit{the lower mass stars seem to host lower mass giant planets}, ranging from $\sim$ 0.5 to 2 $M_J$. As explained in Section \ref{sec:biases}, we do not expect observational biases to account for these planetary mass trends and therefore suggest that on average giant planets are less massive around less massive stars.

\subsubsection{3D: Mass--Radius--Stellar Mass ($M_p, R_p, M_*$)}

To understand this trend further, we expand the dimensionality of this analysis to include planetary radius as the third dimension. We perform a 3D fit to the planetary radius, mass and stellar mass using the cross validation method allowing for a maximum of 200 degrees to pick the optimum number of degrees in each dimension. Based on this we pick 31, 73, and 178 degrees respectively, estimate the weights for each beta function and then calculate the 3D $f(M_p, R_p, M_*)$ joint probability density function (PDF). We then condition this joint distribution at $R_p = 1~R_J$ to focus on warm Jupiters in our sample, i.e., ones that are not hot and inflated beyond 1 $R_J$ \citep{guillot_evolution_2002, miller_heavy-element_2011}. Furthermore, due to the influence of electron degeneracy pressure beyond Saturn mass ($\sim 0.3 M_J$) and Jupiter radius \citep{saumon_theory_1996, 2014prpl.conf..619C}, the planetary radius has a very weak dependence on mass, allowing us to probe for the change in planetary mass with stellar masses.

\begin{figure}[!t]
    \fig{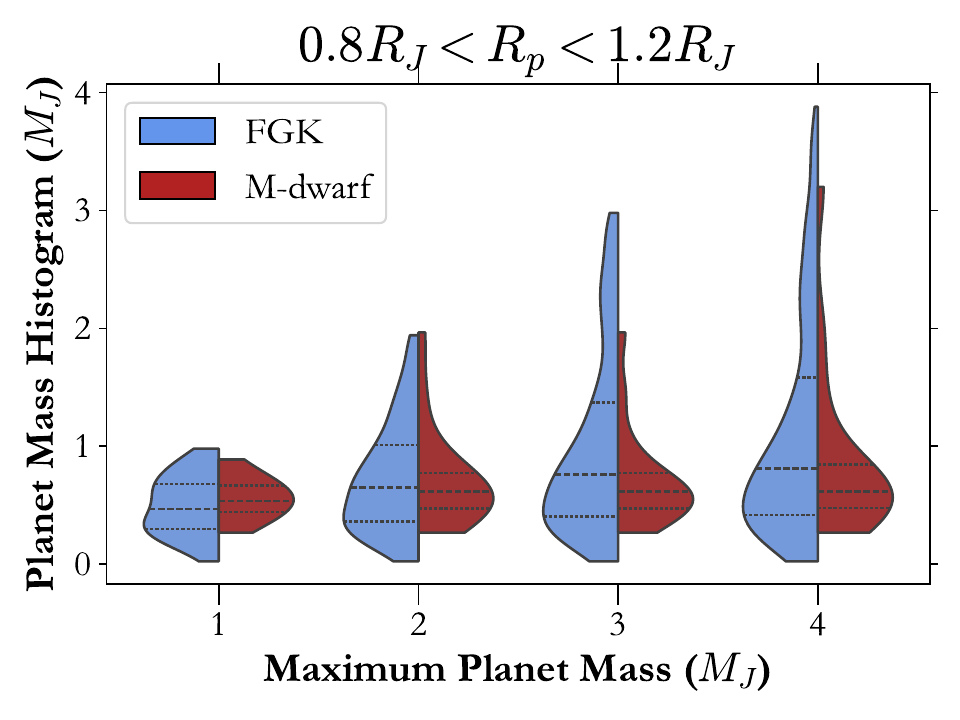}{0.47\textwidth}{}
\caption{\small  Planet mass distributions (Gaussian Kernel-density estimate smoothened histograms) for M-dwarf and FGK samples, when compared across subsamples of planets with masses less than the `Maximum Planet Mass' shown on the x-axis. The vertical lines in each distribution indicate the quartiles. The shape and values of the distributions for the M-dwarf and FGK samples start to converge for planets $\le$ 2 $M_J$. The distributions for the M-dwarf and FGK samples do not seem to have the same mass limits in this plot due to sparsity of the M-dwarf giant planet sample in the planet mass axis (\autoref{fig:scatter}e).}\label{fig:Violin}
\end{figure}

The conditional distributions from our analysis (at $R_p$ = 1 $R_J$) are shown in \autoref{fig:3DPlMassRadius}. The left column includes the full sample of well-characterized transiting Jupiter-sized planets around FGKM stars, and shows that the distributions of planetary masses differ across stellar masses, with a simpler uni-modal distribution for lower-mass stars transitioning to a more complex distribution for more massive stars, which is also seen in the larger spread in planetary masses in \autoref{fig:3DPlMassRadius} (top row) for higher mass stars. Additionally, the expectation value for the planetary mass shows an almost monotonic trend with stellar mass (dashed lines in \autoref{fig:3DPlMassRadius}, bottom row), although given the multi-modal nature of the distribution for higher-mass stars this should be interpreted with caution. Following the methodology outlined in our survey design \citep{kanodia_searching_2024}, we also quantify the dissimilarity in the conditional distribution PDFs --- $f(M_p | R_p = 1 R_J, M_*)$ --- across stellar mass. The Earth-mover distance \citep[EMD;][]{rubner_metric_1998} is a metric to quantify the difference between two distributions without any assumptions of normality or similar variance, and is also known as the Wasserstein distance \citep{kantorovich_mathematical_1960, ramdas_wasserstein_2015}. We bootstrap the sample and estimate the conditional distributions a 100 times to account for finite sample size effects, and find that based on the EMD and Welch's t-test \citep{welch_generalization_1947, ruxton_unequal_2006} we can already reject the null hypothesis that $f(M_p | R_p = 1 R_J)$ is independent of stellar mass, and find that the EMD and T-statistic for $f(M_p | R_p = 1 R_J, M_* = 0.4~$\solmass) is already 3-$\sigma$ discrepant from $f(M_p | R_p = 1 R_J, M_* =$ 1\solmass). This shows that the existing GEMS sample is already enough to perform a point-wise comparison to conclude that the \textit{average} Jupiter radius giant planets around 0.4 \solmass{} stars have different planetary masses (and hence bulk density) than their analogues around solar-type stars. 

However, based on \autoref{fig:scatter}d we see that there are fewer super-Jupiters around lower mass stars. Therefore, to better interpret the multi-modal conditional distributions and directly compare Jupiter-like (in mass and radius) exoplanets, we then filter out super-Jupiters $\ge$ 4 $M_J$ in the middle column of \autoref{fig:3DPlMassRadius} based on the empirical distinction from \cite{santos_observational_2017} and \cite{schlaufman_evidence_2018}. Our analysis of this sample still suggests the presence of a second residual peak in the sample of planets around more massive stars. To check for differences between the distribution of super-Jupiters and Jupiters, we then reduce this cut-off from 4 $M_J$ to 2 $M_J$ (having a cut-off at 1.5 $M_J$ results in a similar conclusion). We see that the masses for Jupiter radius objects are consistent across stellar mass in the right panel of \autoref{fig:3DPlMassRadius}. Using a similar methodology as before across the stellar mass, we cannot reject the null hypothesis that the $f(M_p | R_p = 1 R_J)$ distributions are independent of stellar mass based on the EMD and T-statistic. To ensure that our nonparametric analysis was not smoothing over the small sample of M-dwarf planets and biasing results, we also verify this by performing a categorical comparison between the M-dwarf ($\lesssim$ 4000 K) and FGK sample of giant planets ($\gtrsim$ 4000 K) within 20\% of Jupiter-radius (\autoref{fig:Violin}). The median planet mass for the two samples are within 10\% of each other with a mass cut-off of 2 $M_J$ and below. We also note that given the limited sample size (especially after restricting things based on stellar mass and excluding super-Jupiters), it would be premature to comment on the similarity (or lack there-of) in the shape of the posteriors of the empirical samples (\autoref{fig:Violin}) beyond their median values, and look forward to attempting to draw inferences from shape of these distributions towards the conclusion of our survey.

In other words, despite giant planets spanning masses between 0.3 to 80 $M_J$ at a Jupiter radius due to the radius degeneracy, we show that for Jupiter-radius planets below $<$ 2 $M_J$, their average planet masses are likely independent of host-star mass. 

\subsection{Trends in planet to stellar mass ratio}\label{sec:analysisratios}

\subsubsection{2D: Mass Ratio -- Stellar Mass ($q, M_*$)}
\begin{figure}[!t]
    \fig{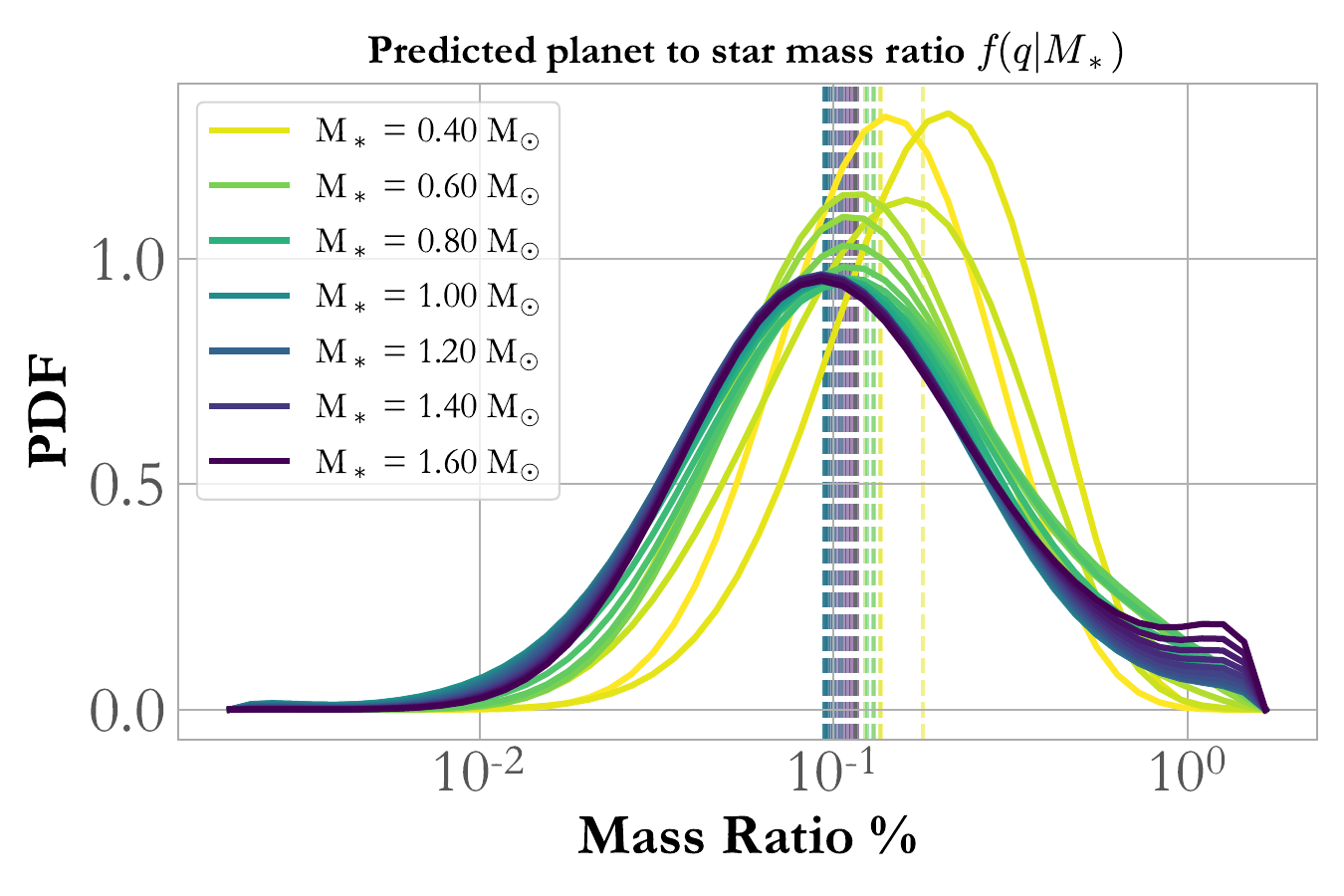}{0.47\textwidth}{}
\caption{\small Similar to \autoref{fig:2D}, but the conditional distribution of planet-to-stellar mass ratio as a function of stellar mass  --- $f(q| M_*)$. This suggests the mass ratio seems to increase towards lower stellar masses. In other words, while the average giant planets around lower mass stars seem to be lower in mass compared to their FGK counterparts, the rate of decrease in planet mass is slower than the decrease in stellar mass.}\label{fig:2DMassRatio}
\end{figure}

We then probe the planet-to-star mass ratio (\textit{q}) as a function of stellar mass  (\autoref{fig:2DMassRatio}), similar to \autoref{fig:2D} and see that for higher mass stars on average the planets have similar mass ratios (0.1\%), i.e., higher mass stars have higher mass planets on average. For lower mass stars, specifically KM stars the average mass ratios are higher. A Kendall-Tau test implemented through \texttt{scipy} \citep{virtanen_scipy_2020} on the GKM sample ($<$ 1.2 \solmass) does indeed find a marginally significant 2\% p-value for a correlation between mass-ratios and stellar mass, suggesting a possible trend. Given the change in the expectation value for mass-ratios across the KM stars, and relative similarity for FGs, it is possible that the decrease in average planet mass with stellar mass for giant planets is not a linear trend, i.e., the rate of decrease in planet mass is slower than the decrease in stellar mass.

\subsubsection{3D: Mass Ratio -- Planet Radius -- Stellar Mass ($q, R_p, M_*$)}

\begin{figure*}[ht]
\centering
\begin{tabular}{ccc}
 \includegraphics[scale=0.55, trim={0.2cm 0.2cm 0.2cm 0.2cm},clip]{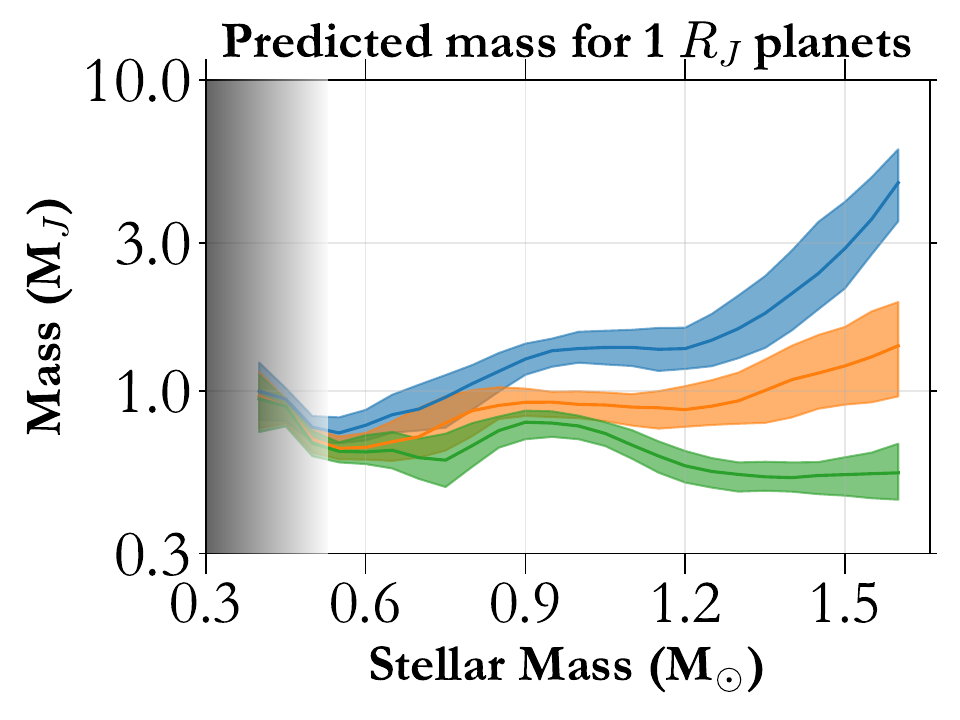} &
 \includegraphics[scale=0.55, trim={0.2cm 0.2cm 0.2cm 0.2cm},clip]{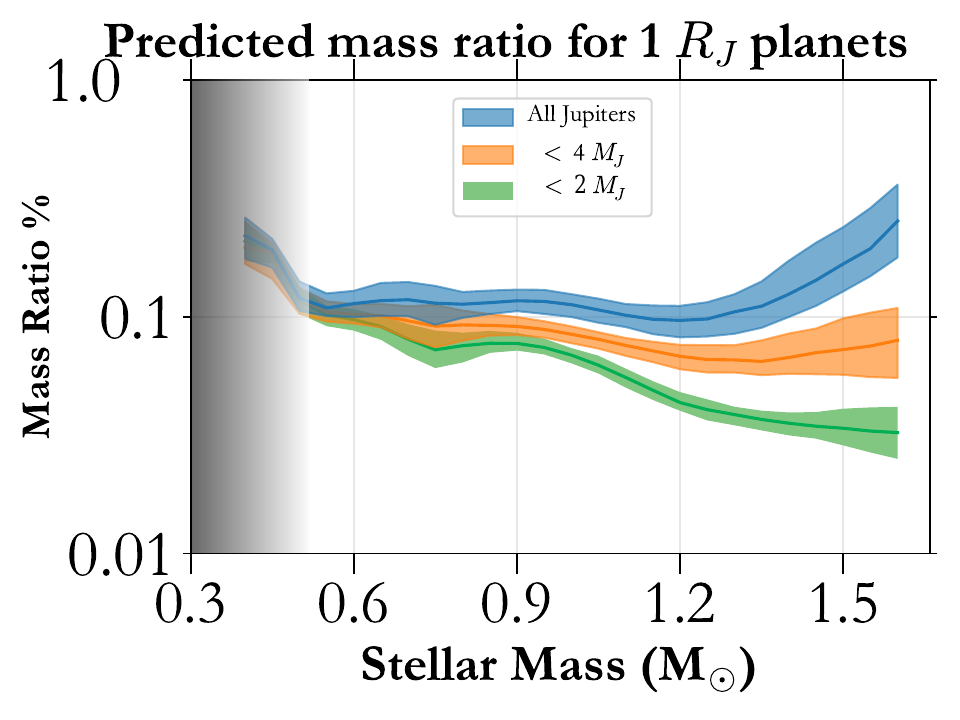} &
 \end{tabular}
\caption{The expectation values for planetary masses (left) and mass ratios (right) as a function of stellar mass. The shaded region shows the 16th--84th percentile distribution of the expectation values from bootstrapping the sample.  Given the few GEMS ($\sim$ 5) around $<$ 0.5 \solmass{} stars (shown in grey), we suggest caution while interpreting trends seen for planets around these stars.}
\label{fig:ExpectationValues}
\end{figure*}

We also investigate the mass ratios with respect to planet radius (to exclude hot Jupiters) and stellar mass and find that, when including all the giant planets (without any super-Jupiter cut), there are similar trends as those explored with planetary masses. We show the expectation value for the planet masses from Section \ref{sec:analysismasses}, and mass ratios (as discussed here) in \autoref{fig:ExpectationValues}.

\section{Results}\label{sec:results}
We draw two main inferences from this analysis, with the caveat that these are focussed on close-in transiting giant planets.

\begin{enumerate}
    \item \textit{When including super-Jupiters, the average Jupiter-sized giant planet mass is dependent on host star mass.} However, this is likely biased by the fact that the prevalence of super-Jupiters seems to be dependent on the host star mass. 
    \item \textit{After accounting for this difference in the prevalence of super-Jupiters by excluding them, we find that the average Jupiter-sized planet masses are independent of stellar mass.} Jupiter-sized planet masses below $\lesssim 2 M_J$ do not seem to show a dependence on stellar mass between 0.4 to 1.6 \solmass. This means that the $\sim$ 4$\times$ difference in stellar mass across our sample (and presumably the formation conditions associated with this) does not result in differences in the mass (or bulk density) for short-period Jupiter-sized exoplanets.
\end{enumerate}

We also note that the M-dwarf spectral sub-type spans a much larger range of stellar masses (0.08 to 0.6 \solmass) than probed by the current GEMS sample. Potential future planets from TESS, as well as astrometric detections from Gaia \citep{perryman_astrometric_2014} will help probe the applicability of the inferences stated above to even lower stellar masses and longer orbital separations, while results from our ongoing \surveyname{} survey will help improve the significance of these results and allow us to probe smaller differences across stellar mass. Furthermore, ongoing atmospheric observations of GEMS with JWST (Cycle 2 GO 3171, 3731, 4227, Cycle 3 GO 5863)  will estimate molecular abundances as well as atmospheric metallicity of these planets to help investigate the presence of similar trends in atmospheric composition.

\begin{figure*}[!t]
    \fig{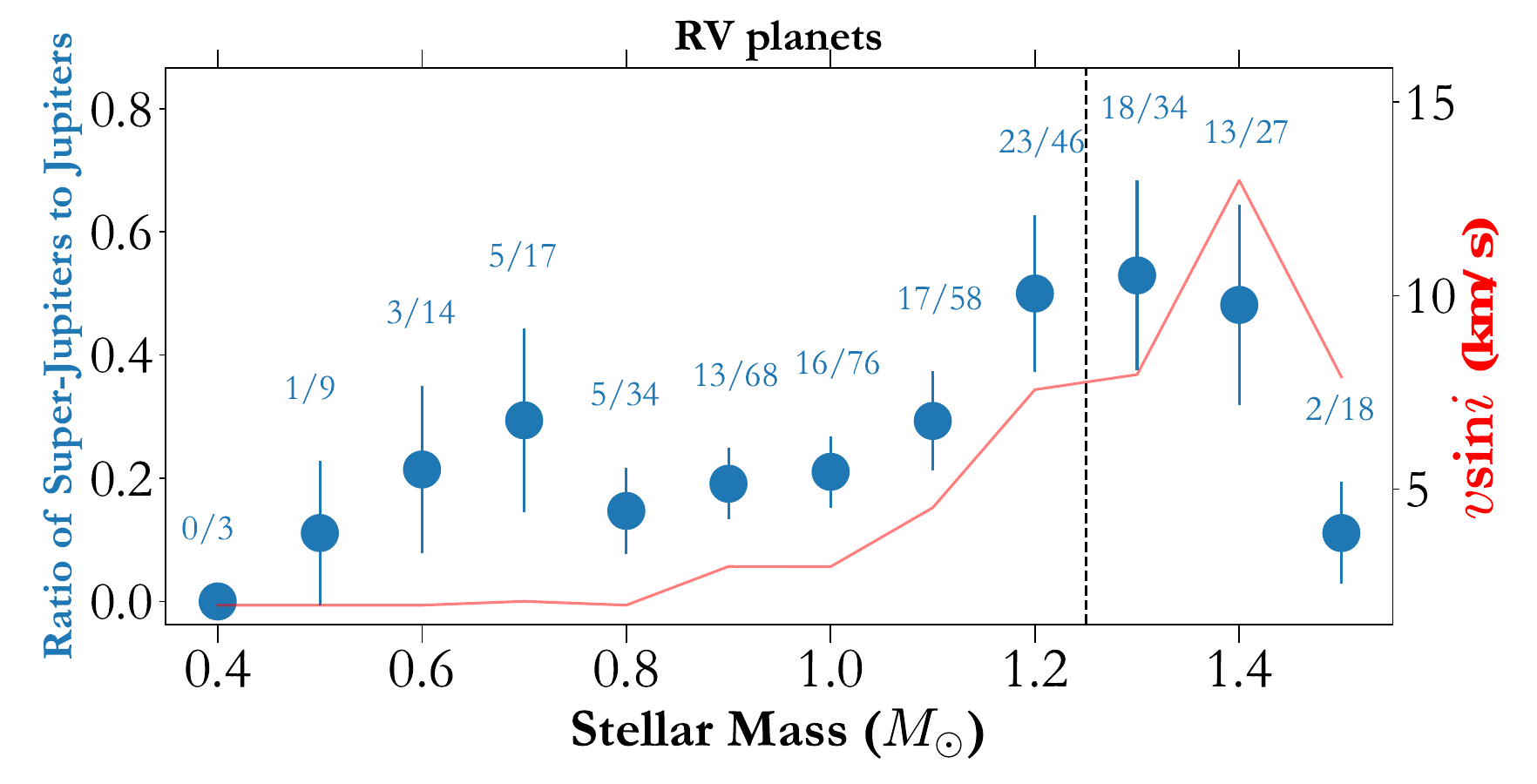}{0.45\textwidth}{}
    \fig{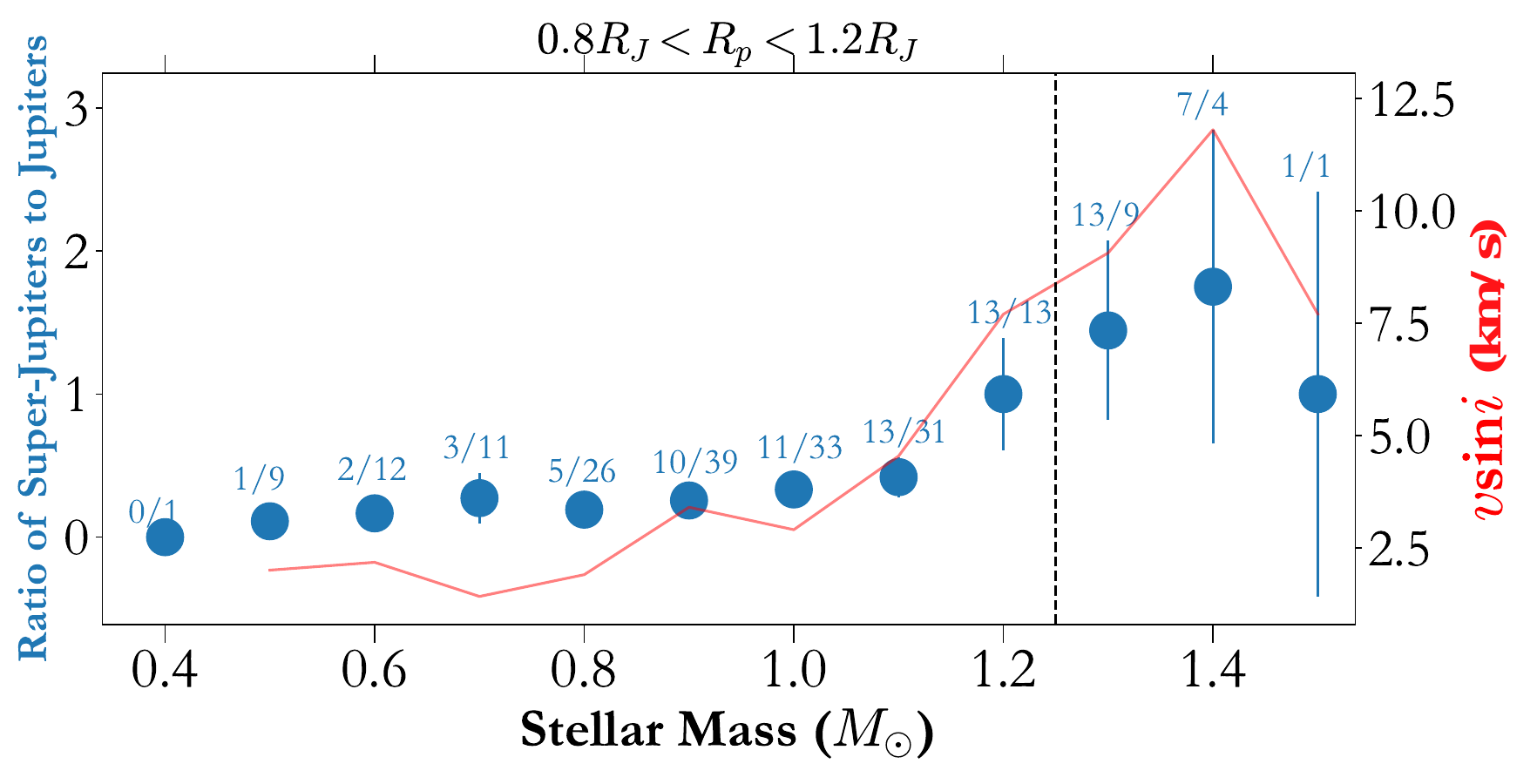}{0.45\textwidth}{}
\caption{\small Ratio of Super-Jupiters to Jupiters as a function of stellar mass with the threshold between the two classes at 2 $M_J$ in blue.  The errors for each bin are propagated after assuming a Poisson distribution, i.e., $\sigma_N$ = $\sqrt N$. The vertical dashed line indicates this tentative transition at 1.3 \solmass, while the numbers in blue indicate the number of Super-Jupiters to Jupiters when using a threshold of 2 $M_J$ between the two. The red line shows the median \vsini{} (projected stellar rotation velocity) for each stellar mass bin. \textbf{Left)} Shows all the giant planets in our sample, where we see the increment in the ratio of super-Jupiters to Jupiters above the Kraft break. \textbf{Right)} We limit our sample to objects close to Jupiter radii (within 20\%) to more closely compare with \autoref{fig:3DPlMassRadius}, and find the relative increase in Jupiter-sized super-Jupiters at the Kraft break much starker.}\label{fig:Ratio}
\end{figure*}

We also speculate on an interesting bi-modality seen in \autoref{fig:3DPlMassRadius} (bottom left), where there seems to be an abrupt transition in the relative number of 1 $R_J$ super-Jupiters to Jupiters at $\sim$ 1.3 \solmass{}, as evinced by the difference in the mode of the conditional distributions. We test this by simply comparing the ratio of the number of super-Jupiters to Jupiters in our sample in 0.1~\solmass{} bins in \autoref{fig:Ratio} (both with and without a radius cut), and find a similar sudden transition in the ratio of super-Jupiters at 1.3 \solmass{}, which incidentally is close to the Kraft break \citep[F5V, $\sim$ 6500 K;][]{kraft_studies_1967, beyer_kraft_2024}. This ratio increases from about 0.38 to 0.71 across the break when considering all giant planets, while focussing on the Jupiter-sized ones (i.e., not inflated) it goes from 0.85 to 1.75.

\section{Discussion}\label{sec:discussion}

\subsection{Data limitation}
We note that the inferences stated above are from only $\sim$ 20 transiting GEMS out of our full FGKM sample of $\sim$ 550 planets. This is especially true for $< 0.5$ \solmass{} stars, with $\sim$ 5 currently known planets. It is therefore clear that a better comparison requires increasing the sample around M-dwarfs. In addition, we need to carefully consider the possible biases in the samples and their heterogeneous nature.
Indeed, the FGK transiting giant planets in our sample are drawn from multiple different surveys and hence have a very heterogeneous selection function. A possible future pathway to overcome this would be to compare masses from warm-Jupiters around FGK stars from homogeneous surveys \citep{dong_warm_2021, yee_tess_2022} with transiting GEMS from our sample. 
As a result, at this stage we can only rely on the available data and speculate the potential implications for giant planet formation of the two inferences stated above.

\subsection{Implications for planet formation}
Our analysis suggests that the average transiting gas giant around an M-dwarf is lower in mass than its FGK counterpart. 
This is largely driven by fewer super-Jupiters around M-dwarfs, which is somewhat expected given that a similar mass planet would have a 2$\times$ mass-ratio around an early M-dwarf, compared to a solar-type star. In order to form such super-Jupiters in the core accretion model, formation in the outer regions ($\sim$ 5 AU) of disks with high-metallicity and high surface density is required.  Only this way the core can grow fast enough to allow for rapid gaseous accretion \citep{mordasini_extrasolar_2009}. 
In the disk instability massive disks that are large and cool enough to initiate instabilities are required \citep{boss_giant_1997, boss_rapid_2006, boss_forming_2023} necessitates. 
In any case, in both of these formation scenarios the formation of massive giant planets around M-dwarfs is extremely challenging.

\par

Interestingly, when we utilize a threshold of $\sim 4 M_J$ to exclude super-Jupiters based on literature studies \citep{santos_observational_2017, schlaufman_evidence_2018}, the above mentioned distinction between stellar mass bins persists. Only after we exclude giant planets $\gtrsim$ 2 $M_J$ from the sample and focus on Jupiters that are a likely product of core-accretion, do we find that the average masses of Jupiter-sized exoplanets become independent of stellar mass. This differs from expectations from core-accretion population synthesis models \citep{ida_toward_2005, alibert_extrasolar_2011, burn_new_2021}. These studies have predicted not just fewer giant planets around M-dwarfs (which is indeed seen in occurrence rate studies), but also lower-mass Jupiters (which is not seen here). This result introduces important constraints on giant planet formation theory.

\begin{figure*}[t!]
\begin{center}
\includegraphics[width=0.75\textwidth]{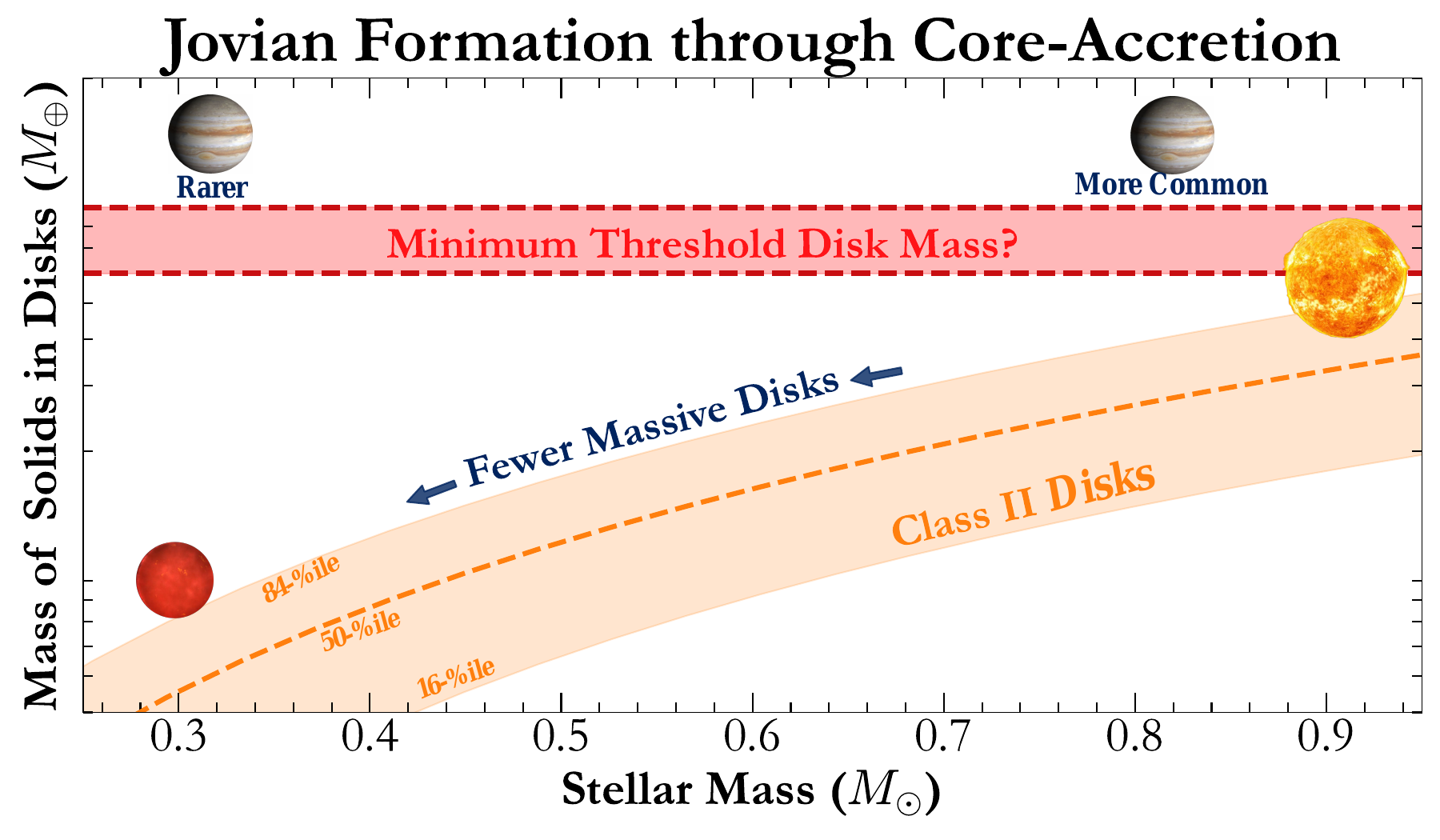}
\end{center}    
\caption{\small A schematic that shows the median ($\pm$ 16--84-th percentile) disk dust masses from the Lupus complex \citep{ansdell_alma_2016} in orange. Our hypothesis for the minimum threshold disk mass is shown in red, where disks more massive than this are able to successfully form Jovians through core-accretion. Since disks greater than this threshold disks mass are rarer around lower mass stars, the occurrence of Jupiters reduces with stellar mass, while the ones that do end up forming have similar masses. We acknowledge the caveat that the disk mass is just one of the many variables that affect giant planet formation. }\label{fig:DiskMass}
\end{figure*}

We hypothesize that the formation of Jupiter-like giant planets necessitates a minimum threshold disk mass, such that they form in similar dust-mass disks across the range of stellar masses studied here (\autoref{fig:DiskMass}). This would imply that while the host star masses decrease, disks forming giant planets remain at similar masses, thereby explaining the decreasing occurrence of giant planets \citep{endl_exploring_2006, johnson_giant_2010, montet_trends_2014, maldonado_connecting_2019, schlecker_rv-detected_2022, pinamonti_hades_2022, gan_occurrence_2023, bryant_occurrence_2023}, as well as the decreasing occurrence of structured disks\footnote{With the assumption being that the gaps and structure in the disks are a product of planet formation.} with stellar mass \citep{van_der_marel_stellar_2021}; since similar (high) mass disks would become rarer around lower mass stars due to the average disk mass to stellar mass scaling. \textbf{In other words, despite the large number of variables that influence giant planet mass, such as the disk mass (also size and surface density), the initial location of formation (pre-migration), and the migration timescale --- which indeed do manifest in the diversity of Jovian planet masses for solar-type stars (Figure \ref{fig:scatter}d) --- the average giant planet masses across stellar masses 0.4 -- 1.6 \solmass{} look similar. }

This result would be expected under the core-accretion paradigm if the final giant planet mass is contingent on the gap that it opens in disks, cutting off additional gas accretion \citep{lin_tidal_1993, bryden_tidally_1999}, with the caveat that the mass that a planet must achieve before it can open up a gap in the disk is also dependent on the stellar mass, disk aspect ratio, semi-major axis, etc. \citep[][]{ginzburg_end_2019}.

Further investigations of giant planet formation around M-dwarfs as well as improved measurements of disks around low-mass stars are required to better understand the observed giant exoplanet populations around different stellar hosts. While here we compare the overall planet mass posterior distribution, larger sample sizes will help put limits on the degree of similarity (i.e., for example, better than 5\%, 10\%, 20\%, etc.). 

We note the caveat that massive disks that are stable around FGK stars, would start to become gravitationally unstable around M-dwarfs due to the lower stellar mass and hence higher disk-to-star mass ratio. That being said, the average Jovians in the current sample (going down to $\sim 0.35$ \solmass) do not necessitate invoking formation in these really massive disks requiring gravitational instability \citep{kanodia_searching_2024, savvidou_there_2024}, which is not necessarily true for the super-Jupiters we exclude from this analysis \citep{delamer_toi-4201_2024}.

We also highlight another salient feature of our study here: for our analysis we picked a minimum radius precision of 5$\sigma$, i.e., 20\%, to limit our analysis to Jovian-sized non-grazing objects and compare them in a mass-degeneracy regime ($\sim$ 1 $R_J$), without confounding inflationary effects seen in hot-Jupiters. However, this also allows the comparisons put forth here to extend to the planetary bulk-density. This is interesting because bulk-density should be a proxy for bulk-metallicity for planets of similar ages and insolation \citep{fortney_planetary_2007, thorngren_mass-metallicity_2016, muller_synthetic_2021}. Differences in inferred bulk-metallicity across stellar masses could point at systematic differences between the ages, stellar metallicity, irradiation, etc. \citep{muller_bulk_2024}, across the sample and should be probed in future work.


Lastly, atmospheric observations from a well-defined sample of seven GEMS spanning the planet mass --atmospheric metallicity plane with the \textit{Red Dwarfs and the Seven Giants: First Insights Into the Atmospheres of Giant Exoplanets around M dwarf Stars} survey \citep[GO 3171;][]{kanodia_red_2023}, as well as other GEMS JWST programs, will provide insights into the atmospheric composition of these planets, thereby enabling comparisons with FGK transiting giant planets. This would extend such comparisons from beyond bulk properties to the atmospheric envelope, which is likely a sensitive tracer of not just formation but also post-formation enrichment in trace quantities \citep{molliere_interpreting_2022, sainsbury-martinez_impact_2024}.

\subsection{Transition at Kraft break?}
Due to rapid increase in the \vsini{} of stars above the Kraft break caused by the disappearance of the outer convective envelope (rendering magnetic braking less effective), it is important to consider a potential observational bias with \vsini{} as the source of this sudden jump in the relative number of super-Jupiters. The radial velocity (RV) information content is inversely proportional to the rotational broadening \citep{bouchy_fundamental_2001, reiners_detecting_2010}, thereby increasing the single-visit RV uncertainty for stars above the Kraft break. 

To ascertain whether this abrupt transition in the ratio of super-Jupiters to Jupiters can be ascribed entirely to the \vsini{} trend, we consider the most conservative scenario and the detectability of not-inflated Jupiters ($<$ 2 $M_J$, $<$ 1.2 $R_J$), i.e., ones farther away from their host star. We estimate the RV uncertainty for the median stellar host (Vmag, \teff, \vsini) at 1.2 \solmass{} (before the break) and 1.3 \solmass{} (after the break), and find a modest increase in the single-visit photon noise estimate from $\sim$ 7 \ms{} to 11 \ms{} in 900 seconds using the NEID Exposure Time Calculator\footnote{\url{https://neid-etc.tuc.noirlab.edu/calc_shell/calculate_rv}} for reference. The RV detection significance ($K/\sigma_K$) across the Kraft break is $>$ 10-$\sigma$, suggesting that these mass measurements are not marginal, and should not be thwarted by the expected increase in RV measurement uncertainty. However, we caution that the median \vsini{} of the observed planet sample is by-definition biased and likely lower than that from a control sample of F-type stars.

Therefore, out of an abundance of caution, the heterogeneous nature of our sample, and the lack of published planet mass-based selection functions for transit surveys, we consider the purported abrupt increase in the number of super-Jupiters above the Kraft break to be tentative. To test this, we considered searching for similar trends in samples from RV surveys, however the sin$i$ degeneracy and subsequent stochasticity obfuscates a clear delineation between Jupiters and super-Jupiters. The best test to test the validity of this purported trend would be to obtain planet masses for a controlled sample of transiting warm-Jupiters around F-type stars on both sides of the Kraft break.

\section{Conclusion}\label{sec:conclusion}

We compare the bulk properties of transiting giant planets ($\gtrsim$ 8\earthradius) spanning a range from stellar masses ranging from mid M-dwarfs to F-type stars (\autoref{fig:scatter}). With just about 2 dozen transiting GEMS (out of the sample of $\sim$ 550), we are already starting to see interesting trends in the data, which will be tested further with the ongoing \surveyname{} as well as community detections and confirmations of such planets enabled by the TESS mission. 

We make the following inferences with respect to transiting giant planets: 1) Driven by fewer super-Jupiters ( $\gtrsim$ 2 $M_J$) around M-dwarfs, the average M-dwarf Jupiter is lower in mass than its analogue around solar-type stars (\autoref{fig:2D}), a trend that persists while comparing with only FGK warm-Jupiters (\autoref{fig:3DPlMassRadius}, bottom left). 2) After excluding super-Jupiters from the sample, we see that the average M-dwarf Jupiter is similar in mass to the FGK warm-Jupiters (\autoref{fig:3DPlMassRadius} bottom right, \autoref{fig:Violin}, \autoref{fig:ExpectationValues}). We also see trends consistent with these conclusions with planet-to-star mass ratio (\autoref{fig:2DMassRatio}). We hypothesize that these trends could be a result of a minimum disk dust mass threshold that is required for Jovian formation through core-accretion (\autoref{fig:DiskMass}), which explains the rarer occurrence of such planets around M-dwarfs, and also their similarity in mass with FGK Jupiters. Furthermore, ongoing and future atmospheric observations will enable similar comparisons of the atmospheres of these planets across the stellar mass axis, as a probe of their post-formation enrichment. 

Lastly, we speculate on an abrupt transition in the ratio of super-Jupiters to Jupiters for F-type stars at the Kraft break (\autoref{fig:Ratio}), which could be a detection bias of increased \vsini{} above the break for stars. A targeted survey of giant planet candidates around F-type stars, with published non-detections, should help test the validity of this trend. Overall, our results provide valuable insights into the formation and evolution of giant exoplanets across a diverse range of stellar environments.

\section{Acknowledgements}

I thank the referee for their suggestions and comments which have helped clarify the message of this manuscript. I would also like to acknowledge discussions with Ravit Helled and Suvrath Mahadevan that helped mould the science narrative for this paper, and also feedback from Johanna Teske, Caleb Canas, Alycia Weinberger, John Chambers, Peter Gao, Alan Boss, Nicole Wallack and Dana Anderson that have helped expand the scope of this paper, while crystallizing its message. I also acknowledge help from Theodora for proof-reading this manuscript.

I also thank Peter Gao for help with the computing resources that enabled running some of these memory intensive analyses. 

This research made use of the (i) NASA Exoplanet Archive, which is operated by Caltech, under contract with NASA under the Exoplanet Exploration Program, (ii) SIMBAD database, operated at CDS, Strasbourg, France, and (iii) NASA's Astrophysics Data System Bibliographic Services.

\software{
\texttt{astropy} \citep{robitaille_astropy:_2013, astropy_collaboration_astropy_2018},
\texttt{ipython} \citep{perez_ipython:_2007},
\texttt{matplotlib} \citep{hunter_matplotlib:_2007},
\texttt{MRExo} \citep{kanodia_mass-radius_2019, kanodia_beyond_2023},
\texttt{numpy} \citep{oliphant_numpy:_2006, harris_array_2020},
\texttt{pandas} \citep{mckinney_data_2010},
\texttt{scipy} \citep{oliphant_python_2007, virtanen_scipy_2020},
}

\bibliography{MyLibrary,OtherReferences}

\end{document}